\documentclass[conference]{IEEEtran}
\IEEEoverridecommandlockouts

\usepackage{cite}
\usepackage{booktabs}
\usepackage[normalem]{ulem}
\usepackage{amsmath,amssymb,amsfonts}
\usepackage{algorithmic}
\usepackage{graphicx}
\usepackage{svg}
\usepackage{textcomp}
\usepackage{xcolor}
\usepackage{hyperref}
\def\BibTeX{{\rm B\kern-.05em{\sc i\kern-.025em b}\kern-.08em
    T\kern-.1667em\lower.7ex\hbox{E}\kern-.125emX}}
\begin{document}

\title{Toward Explainable NILM: Real-Time Event-Based NILM Framework for High-Frequency Data\\
% {\footnotesize \textsuperscript{*}Note: Sub-titles are not captured for https://ieeexplore.ieee.org  and
% should not be used}
% \thanks{Identify applicable funding agency here. If none, delete this.}
}

% \author{\IEEEauthorblockN{1\textsuperscript{st} Given Name Surname}
% \IEEEauthorblockA{\textit{dept. name of organization (of Aff.)} \\
% \textit{name of organization (of Aff.)}\\
% City, Country \\
% email address or ORCID}
% \and
% \IEEEauthorblockN{2\textsuperscript{nd} Given Name Surname}
% \IEEEauthorblockA{\textit{dept. name of organization (of Aff.)} \\
% \textit{name of organization (of Aff.)}\\
% City, Country \\
% email address or ORCID}
% \and
% \IEEEauthorblockN{3\textsuperscript{rd} Given Name Surname}
% \IEEEauthorblockA{\textit{dept. name of organization (of Aff.)} \\
% \textit{name of organization (of Aff.)}\\
% City, Country \\
% email address or ORCID}
% \and
% \IEEEauthorblockN{4\textsuperscript{th} Given Name Surname}
% \IEEEauthorblockA{\textit{dept. name of organization (of Aff.)} \\
% \textit{name of organization (of Aff.)}\\
% City, Country \\
% email address or ORCID}
% \and
% \IEEEauthorblockN{5\textsuperscript{th} Given Name Surname}
% \IEEEauthorblockA{\textit{dept. name of organization (of Aff.)} \\
% \textit{name of organization (of Aff.)}\\
% City, Country \\
% email address or ORCID}
% \and
% \IEEEauthorblockN{6\textsuperscript{th} Given Name Surname}
% \IEEEauthorblockA{\textit{dept. name of organization (of Aff.)} \\
% \textit{name of organization (of Aff.)}\\
% City, Country \\
% email address or ORCID}
% }

\makeatletter
\newcommand{\linebreakand}{%
  \end{@IEEEauthorhalign}
  \hfill\mbox{}\par
  \mbox{}\hfill\begin{@IEEEauthorhalign}
}
\makeatother

\author{
\IEEEauthorblockN{Grigorii Gerasimov\IEEEauthorrefmark{1}, Ilia Kamyshev\IEEEauthorrefmark{1}\IEEEauthorrefmark{2}, Sahar Moghimian Hoosh\IEEEauthorrefmark{1}\IEEEauthorrefmark{2}, Elena Gryazina\IEEEauthorrefmark{1}, Henni Ouerdane\IEEEauthorrefmark{1}}
\IEEEauthorblockA{\IEEEauthorrefmark{1}\textit{Skolkovo Institute of Science and Technology}, Moscow, Russia}
\IEEEauthorblockA{\IEEEauthorrefmark{2}\textit{Monisensa Development LLC.}, Moscow, Russia}
\IEEEauthorblockA{Email: Ilia.Kamyshev@skoltech.ru}
}

\maketitle

\begin{abstract}

Non-Intrusive Load Monitoring (NILM) is an advanced, and cost-effective technique for monitoring appliance-level energy consumption. However, its adaptability is hindered by the lack of transparency and explainability. To address this challenge, this paper presents an explainable, real-time, event-based NILM framework specifically designed for high-frequency datasets. The proposed framework ensures transparency at every stage by integrating a z-score-based event detector, appliance signature estimation, Fourier-based feature extraction, an XGBoost classifier, and post hoc SHAP analysis. The SHAP analysis further quantifies the contribution of individual features, such as cosine of specific harmonic phases, to appliance classification. The framework is trained and evaluated on the PLAID dataset, and achieved a classification accuracy of 90\% while maintaining low computational requirements and a latency of less than one second. 

\end{abstract}


\section{Introduction}

As energy consumption continues to grow annually, load monitoring has become important. In the European Union, buildings account for approximately 41\% of electricity consumption \cite{González_2022}. This underscores the need for advanced energy management solutions in the building sector which enable effective monitoring, control, and optimization of electricity. 
It is known that providing real-time feedback on energy usage can lead to significant savings \cite{Darby_2006}: Direct feedback, such as smart meters and real-time energy display monitors, can reduce consumption by 5\% to 15\%; indirect feedback, including detailed utility bills and periodic reports, may result in savings of up to 10\%. Both methods enhance user awareness and encourage energy-efficient behavior.

Load monitoring methods are categorized as intrusive load monitoring (ILM) and non-intrusive load monitoring (NILM) \cite{Tokam_2023}. ILM involves installing sensors or sub-meters on individual appliances. This method provides high accuracy, but it is expensive, difficult to install, and unsuitable for large-scale applications. NILM, on the other hand, monitors the total energy consumption at a single point, typically the main electrical panel, and uses disaggregation algorithms to estimate the energy usage of individual appliances. NILM is cheaper, easier to deploy, and more scalable than ILM. However, its accuracy depends on the quality of the load disaggregation algorithms it uses. 
Much research devoted to NILM aims to improve the accuracy and efficiency of load disaggregation. A detailed review of NILM methods is provided in \cite{Angelis_2022, Rafiq_2024}.

\begin{figure*}[t]
\centering
\includegraphics[width=\textwidth]{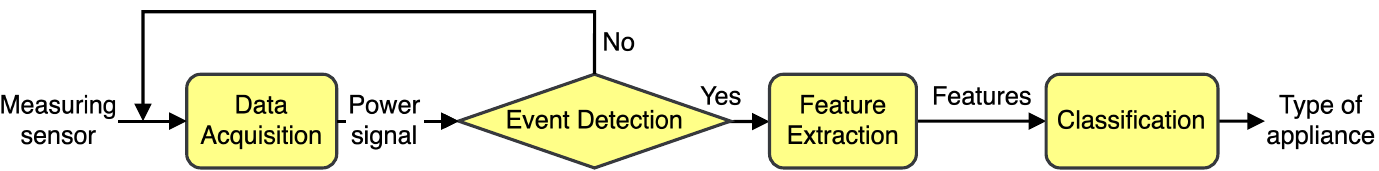}
\caption{Block diagram of an event-based NILM system.}
\label{event_based_NILM_algo}
\end{figure*}

NILM approaches can be broadly classified into event-based and non-event-based approaches \cite{Anderson_2012}. Event-based methods detect significant changes in power consumption, such as when appliances are turned on/off. These methods are efficient and suitable for real-time monitoring. However, they may face challenges under noisy conditions or when multiple appliances change states at the same time. Non-event-based methods continuously analyze power consumption data without explicitly detecting events. Although non-event-based methods are more accurate in complex scenarios, they require more computational resources and large labeled datasets for training \cite{Anderson_2012}.

In recent years, both event-based and non-event-based NILM models have increasingly adopted machine learning (ML) and deep learning (DL) techniques due to their powerful ability to capture complex and non-linear patterns in data. These models, especially DL, are highly effective at finding intricate relationships and learning detailed features directly from raw data. However, the features learned by DL models are often abstract and not easily interpretable, posing challenges for researchers and users to understand and trust NILM models \cite{machlev2022explainable}. One of the major criticisms of DL models is their "black box" nature, as they often provide high accuracy but little insight into their decision-making processes. Explainable AI (XAI) is a promising approach to enhance the transparency in the decision-making process of AI systems. XAI methods include (I) transparent models, such as decision trees and regression models, which are inherently interpretable in case if features are not multi-collinear, and (II) post hoc techniques that explain predictions.

In the context of NILM, XAI is an emerging area, with only a few studies discussing it. For instance, early work by Murray et al. \cite{Murray_2020} proposed masking appliance activations, i.e., intentionally removing an appliance from input data and then analyzing its effect on the inner working process of a one-dimensional autoencoder model. Although insightful, this approach was limited to a single appliance and lacked scalability. Another work \cite{machlev2022explaining} explored explainability for convolutional neural networks (CNNs) in NILM, employing post hoc techniques like Occlusion sensitivity and gradient class activation mapping on the low-frequency REDD dataset (sampling rate at 1 Hz). Then, the important features were visualized using heatmaps, which can be challenging for non-expert users to interpret. Furthermore, work \cite{batic2023towards} applied advanced explanation techniques to the regression-based DL NILM using datasets like UK-DALE and REFIT (both low-frequency data). They introduced metrics such as faithfulness, robustness, and complexity for evaluating explainability but noted the need for standardized benchmarks and scalable solutions for large datasets.

All these works highlight the growing importance of transparency and explainability in NILM, while also revealing gaps in computational efficiency, scalability, and simplicity. Moreover, for a truly transparent NILM approach, explainability must extend beyond the modeling stage. This requires making the entire framework transparent, including preprocessing, event detection, and feature extraction. A transparent framework for NILM addresses these needs by ensuring that every stage of the process is self-explanatory. Such a framework can benefit all stakeholders as it improves reproducibility for researchers, delivers actionable insights to end-users, and helps engineers interpret results. 

In this paper, we address the need for trustworthy NILM models by introducing an explainable framework for event-based NILM. To the best of our knowledge, this is the first explainable framework specifically designed for high-frequency datasets (in the range of kHz). Our end-to-end framework enables real-time disaggregation with a minimum latency of 320/380 ms for 60/50 Hz power system, respectively. In this work, we aim to build trust in NILM systems and promote their adoption in real-world applications.

\subsection{Contributions}

 \begin{itemize}
     \item We present the first explainable end-to-end and real-time NILM framework for high-frequency data.
     
    \item The proposed method is suitable for real-time applications on edge devices with a latency of $19/f_0+\tau$ seconds, where $f_0$ is a fundamental frequency and $\tau$ is an execution time.

    \item We select a small set of least correlated and interpretable features of appliance signatures that result in 90\% classification accuracy on real-world dataset.

    \item We make the source code for the proposed framework publicly-available on GitHub:
    \href{https://github.com/arx7ti/xai-nilm}{https://github.com/arx7ti/xai-nilm}

 \end{itemize}

The rest of the paper is organized as follows: Section~\ref{back} provides the necessary background on the key components of event-based NILM, and reviews key approaches. Section~\ref{method} provides the implementation details of the proposed framework. Section~\ref{results} presents experimental result on the PLAID dataset. Section~\ref{discussion} provides a latency and complexity analysis for the proposed framework. Finally, Section~\ref{conclusion} concludes the study, and suggests directions for future research.

\section{Background}
\label{back}
The concept of event-based NILM was first introduced by Hart \cite{Hart_1992}. He proposed a method for identifying appliances by detecting significant changes in active and reactive power, referred to as events. This approach works well for appliances with distinct on/off states but encounters challenges with overlapping events or variable loads. Figure \ref{event_based_NILM_algo} illustrates the main stages of the event-based NILM framework, which will be discussed further in detail. 

\subsection{Data acquisition}

Data acquisition is the initial step, where measuring sensor continuously record power consumption data. Key parameters include voltage, current, active power and reactive power \cite{Rafiq_2024}. These measurements form the foundation for event detection and feature extraction. In particular, high-frequency sampling signals capture maximal information about activated/deactivated appliance due to harmonic content \cite{Angelis_2022}.

\subsection{Event detection}

Event detection identifies step changes in power signals caused by appliances turning on/off. These changes can be detected by using rule-based, probabilistic methods or other approaches. Probabilistic models achieve detection rates up to 95\% \cite{Alcalá_2017}, and often outperform heuristic methods. Recent hybrid methods, such as those combining probabilistic and heuristic approaches \cite{Zhang_2022}, improve robustness and accuracy. For example, voting improved isolated forest (VIIF) and time shift downsampling matching (TSDM) are used for predetection and verification of appliance events. However, these methods are frequently not very simple to interpret due to their reliance on complex mathematical formulas.

\subsection{Feature extraction}

Once an event is detected, the next step is to extract relevant features that will help in appliance classification. A review on the feature selection for NILM is provided in \cite{features}. In high-frequency NILM, common features include the harmonic amplitudes and phases, wavelets, voltage-current (V-I) trajectories, current over time, phase shift etc. Study \cite{features} reports that there is no universal combination of features as each model behaves differently. However, it is claimed that the best features always reflect spectral information.

\subsection{Classification}

The next step is to classify the appliances based on features computed previously. The classification process employs various ML/DL approaches such as convolutional neural networks, decision trees, gradient boosting models etc. Neural networks often result in higher classification accuracy at the high cost of interpretability due to the "black-box" nature of the networks. One of the examples of promising event-based neural network architectures is concatenate-CNN \cite{Wu_2019}. Gradient boosting models have higher interpretability of predictions as they are built of multiple decision trees \cite{Himeur_2020}, and offer feature importance measure. In its turn, decision trees are typically considered as fully transparent machine learning models in case if there is no pair of multi-collinear features.

\begin{figure*}[t!]
\centering
\includegraphics[width=0.8\textwidth]{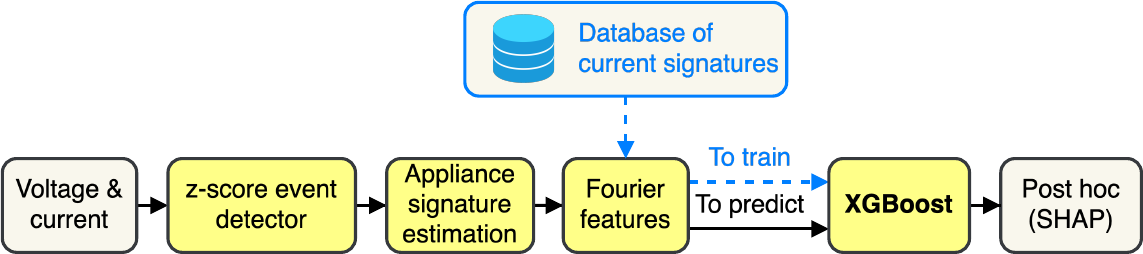}
\caption{The flow chart of a proposed approach for high-frequency event-based NILM.}
\label{our_approach}
\end{figure*}

\section{Proposed method}
\label{method}

\begin{figure*}[t]
\centering
 \includegraphics[width=\textwidth]{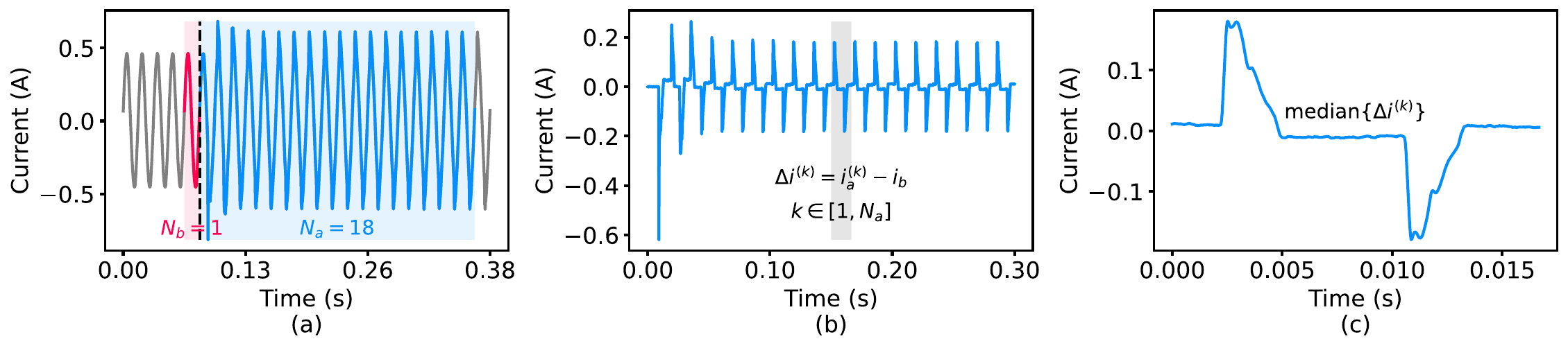}
 \caption{Activation current processing for a turn-on event. 
 (a) Aggregated current signal with highlighted $N_b = 1$ cycle before and $N_a = 18$ cycles after an event. 
 (b) Activation current $\Delta i$ calculated for 18 cycles.
 (c) Estimated appliance current signature calculated as in Eq.~\ref{eq:est}.}
\label{fig:before-after}
\end{figure*}

In this section, we propose a framework for high-frequency event-based NILM which is suitable for real-time applications on edge devices. Each stage of our approach has a clear and relatively simple mathematical foundation that significantly increases overall transparency compared to other methods. Figure \ref{our_approach} illustrates the flow chart of the proposed approach. 

The main goal of the framework is to detect and classify on/off events of appliances and explain the results. After detecting an event, an appliance signature is estimated based on the activation/deactivation current defined in \cite{Baets_2018, Faustine_2020_act}. Later on, eight Fourier features are extracted from the obtained signature and passed to the XGBoost model (a gradient boosting model). Finally, post hoc analysis is applied to explain why a particular device was classified as on/off.

\subsection{Real-time voltage and current processing}

At the first stage, the proposed system records cycles of voltage (\textit{v}) and current (\textit{i}). Given the fact that the grid's frequency is prone to fluctuations, the frequency-invariant transformation of periodic signals (FIT-PS) is applied to both recorded voltage and current signals. This step is essential for simplifying further mathematical operations. The FIT-PS algorithm detects zero-crossings of the voltage signal and resamples both voltage ($\tilde{v}$) and current ($\tilde{i}$) vectors to a common length $T$ ($T= 500$ for 30 kHz data), as described in \cite{Held_2018}:

\begin{equation}
    \tilde{v},\tilde{i}=\text{FITPS}(v, i)
\end{equation}

\subsection{z-score event detector}

Once the voltage ($\tilde{v}^{(k)}$) and current ($\tilde{i}^{(k)}$) signals of $k$-th cycle are obtained, the active power $p^{(k)}$ is computed as:

\begin{equation}
    p^{(k)} = \frac{1}{T} \sum_{t=1}^T\tilde{v}_t^{(k)} \cdot \tilde{i}_t^{(k)}
\end{equation}

The z-score event detector operates on the each new active power computed. An event is detected when the z-score ($z^{(k)}$) exceeds a predefined threshold $Z$ (we set $Z=30$), indicating a statistically significant change in power at the new observed cycle $k$:

\begin{equation}
    z^{(k)} = \frac{|p^{(k)} - \mu_w|}{\sigma_w}, \quad z^{(k)} > Z,
\end{equation}

where $\mu_w$ and $\sigma_w$ are the mean and standard deviation, respectively. The index $w$ is related to the size of a sliding window of previous $w$ cycles prior to the cycle $k$ (in our case $w=10$). To reduce false positives caused by rapid transient behavior, a \textit{window reset} is applied after the each detected event. That is, a statistic should be collected from scratch for the next $w$ cycles, causing a so-called blind zone.

\subsection{Appliance signature estimation}

After detecting an event, the system estimates a power signature of an appliance that was turned on/off via computing an activation/deactivation current. The activation/deactivation current was first introduced in \cite{Baets_2018} and defined as the difference between the current after an event and the current before an event.

The authors \cite{Faustine_2020_act} considered ten cycles before and two cycles after an event to compute the activation/deactivation current. In this work, we consider only one cycle before the event and 18 cycles after the event, as illustrated in Fig.~\ref{fig:before-after}(a-b). Figure ~\ref{number_cycles} suggests that choosing 18 cycles after the event ensures the best accuracy for the classification model. Therefore, at $N_a = 18$, the accuracy reaches its maximum value and does not improve significantly with more cycles. Finally, the activation current, $\Delta i$, is computed for each cycle $k$ as follows:

\begin{equation}
    \Delta i^{(k)} = i_a^{(k)} - i_b,
\end{equation}

where $i_a^{(k)}$ is a vector of instantaneous values of current after an event. Similarly, a vector $i_b$ represents a cycle before the event.

Finally, we extract median cycle to obtain an estimation of an appliance signature ($i_{est}$) which is robust to noise and to transient process, see Fig.~\ref{fig:before-after}(c):

\begin{equation}
\label{eq:est}
    i_{est}=\text{median}\{\Delta i^{(k)}\}
\end{equation}

\begin{figure}[b!]
\centering
 \includegraphics[width=0.75\columnwidth]{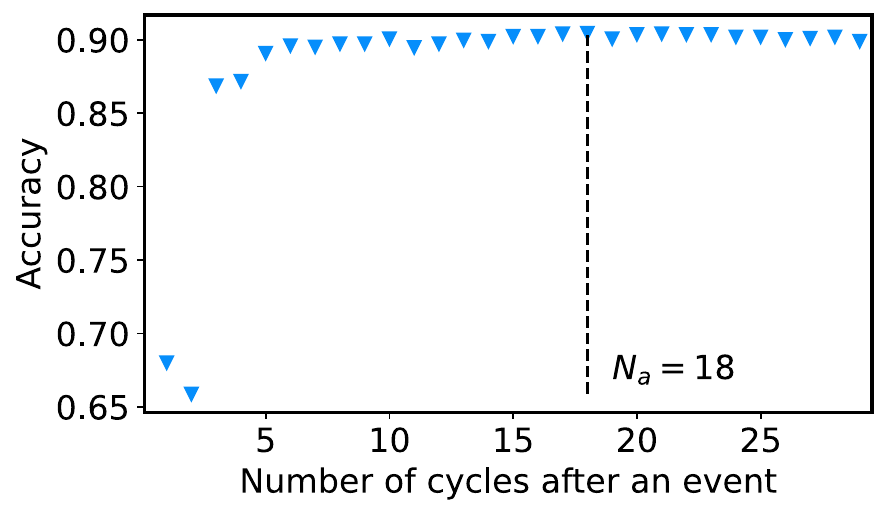}
\caption{Impact of the number of cycles after an event on the appliance classification accuracy.}
\label{number_cycles}
\end{figure}

\begin{figure}[t]
\centering
 \includegraphics[width=0.75\columnwidth]{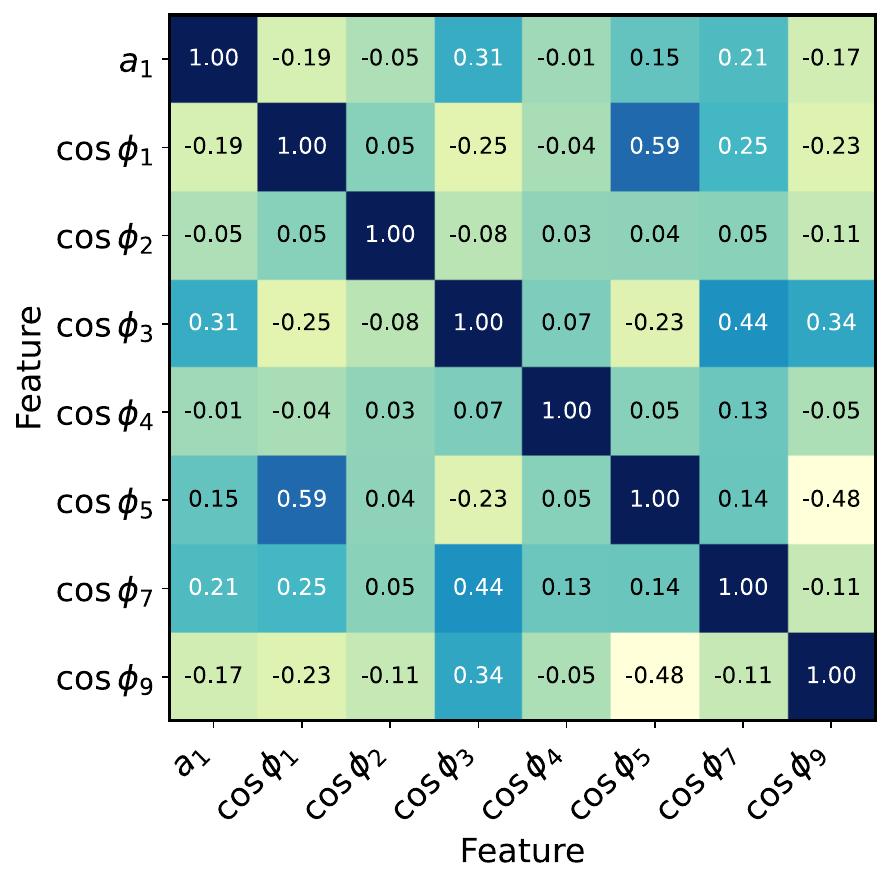}
\caption{Correlation matrix of Fourier features.}
\label{correlation_matrix}
\end{figure}

\subsection{Fourier features}

The estimated current signature ($i_{est}$) is processed using the discrete fast Fourier transform (DFFT) to extract spectral features that characterize appliance behavior. We examined the impact of the full range of harmonics on the model output and observed that the amplitudes ($a_i$) of the second- and higher-order harmonics are strongly correlated, whereas the corresponding phases ($\phi_i$) show weaker correlations. Additionally, we determined that only ten phases have the highest influence on the model's output, based on feature importance analysis performed after fitting the gradient boosting classification model.

However, notion of low and high values of phase does not provide provide meaningful information for users due to the cyclic property. To enhance interpretability, we applied the cosine function to each phase $\phi_i$. Thus, transformed phase features have values in the range [-1, 1]. Finally, we propose the following set of eight features that characterize appliances: $\{a_1, \cos\phi_1, \cos\phi_2, \cos\phi_3, \cos\phi_4, \cos\phi_5, \cos\phi_7, \cos\phi_9\}$.

As seen in the feature correlation matrix in Fig.~\ref{correlation_matrix}, the selected features have weak correlations, which can increase the potential for explainability. In fact, the lower the correlation between features, the more unique information each feature contributes to the decision-making process. Hence, it is more likely to understand which feature impacted the prediction of one or another type of appliance.

\subsection{XGBoost, a gradient boosting model}

As was discussed above, a gradient boosting model offers a good balance between model performance and interpretability. In this work, eXtreme Gradient Boosting (XGBoost) is used to classify types of activated/deactivated appliances.

The model is trained on one-cycle appliance signatures extracted from the reference dataset. From these, eight features are computed as previously described. At the inference stage, the estimated appliance signature is passed to the model for classification.

To find the best hyperparameters of XGBoost model, a random search was performed. As a result, the model has 150 estimators, maximum depth of 8, learning rate of 0.046, and a regularization coefficient $\alpha=10$.

\subsection{Post hoc (SHAP)}

\begin{figure}[t!]
\centering
 \includegraphics[width=\columnwidth]{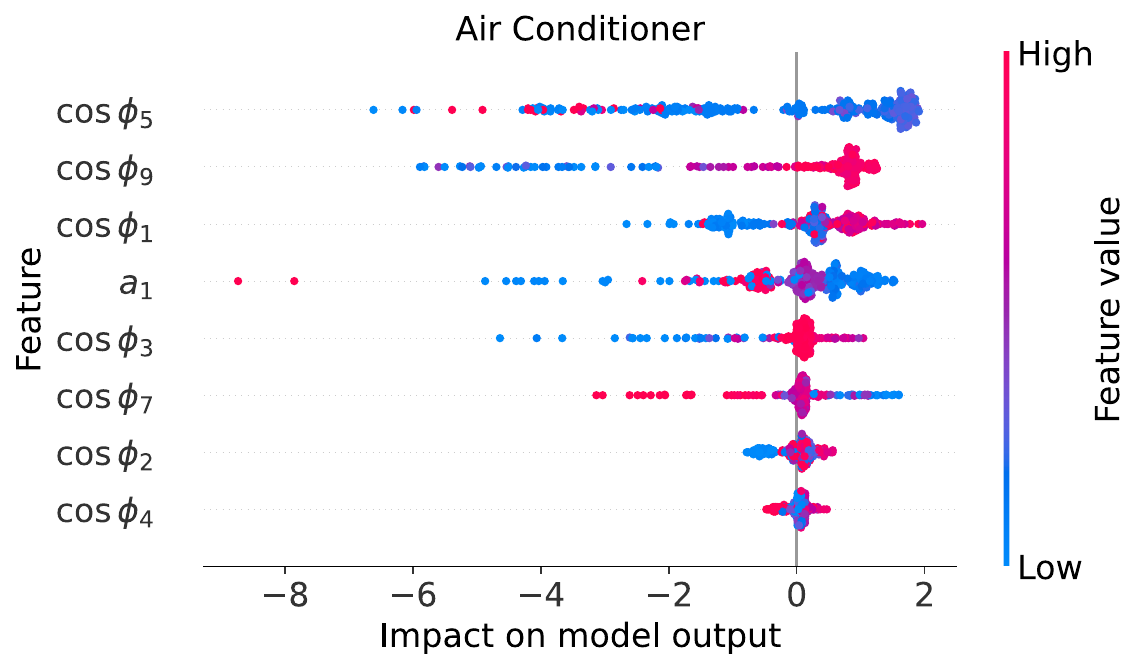}
\caption{SHAP summary plot for features contributions to XGBoost predictions.}
\label{shap_summary}
\end{figure}

A post hoc analysis is extremely important for increasing the trust in NILM algorithms. To ensure the explainability of predictions obtained with the use of our framework, SHapley Additive exPlanations (SHAP) is applied. SHAP is a model-agnostic tool that quantifies the contribution of each individual feature to a model's prediction. Its reliability and effectiveness are widely recognized in research community, e.g., \cite{machlev2022explainable} highlighted that SHAP is one of the most trusted and extensively used XAI techniques.

Figure~\ref{shap_summary} visualizes the contribution of each of the eight features to predicting the air conditioner as an activated/deactivated appliance. It highlights the interpretability of cosine features in correctly classifying air conditioners as true positive or true negative classes. For example, low cosine values of phases for harmonics 1, 3, 5, and 9 indicate that the signature unlikely corresponds to an air conditioner. In contrast, low values of $\cos\phi_7$ indicate that the signature corresponds to an air conditioner. Medium to high values imply a stronger likelihood that the signature belongs to the air conditioner class. High values of the fundamental harmonic amplitude have minimal impact on the model's output. 

\section{Experimental results}
\label{results}
\subsection{Data preparation}

The proposed method was evaluated using the aggregated data of PLAID dataset \cite{Medico_2020}, a publicly available dataset for NILM research. PLAID contains 575 high-frequency voltage and current measurements recorded at a sampling rate of 30 kHz, and annotated on/off events for 13 types of appliances. Our finding reveals there are at least 11 different brands of devices in total that were annotated.

In this work, the dataset was divided into train and test subsets. For training, 164,510 single cycle current signatures of individual devices were extracted. These signatures form a database indicated in Fig.~\ref{our_approach}. For testing, 2,347 activation/deactivation currents were obtained by using z-score detector and Eq.~\ref{eq:est}. The test labels were assigned as a device type that was turned on/off after an event. Thus, the train dataset comprises the original individual signatures of appliances and their respective labels, and the test dataset comprises the estimated current signatures and their labels.

\subsection{Appliance classification performance}

\begin{figure}[t]
\centering
 \includegraphics[width=\columnwidth]{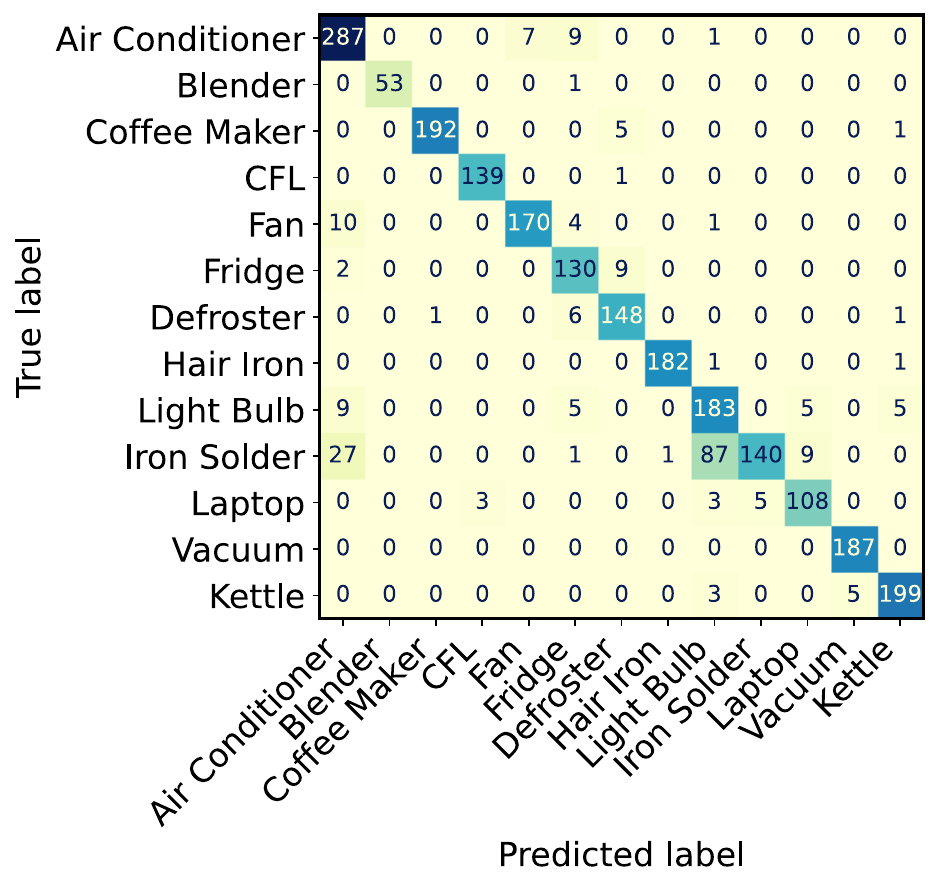}
\caption{Confusion matrix for appliance classification using the XGBoost model.}
\label{confusion_matrix}
\end{figure}

\begin{table}[tbp]
\caption{Performance metrics for classification models}
\begin{center}
\begin{tabular}{ccccc}
\toprule
\textbf{\textit{Model}} & \textbf{\textit{Accuracy}} & \textbf{\textit{Precision}} & \textbf{\textit{Recall}} & \textbf{\textit{F1}} \\ \midrule
XGBoost & \textbf{90\%} & \textbf{92\%} & \textbf{92\%} & \textbf{91\%} \\
\hline
Decision Tree & 84\% & 85\% & 86\% & 84\% \\
\hline
Logistic Regression & 80\% & 80\% & 80\% & 80\% \\
\bottomrule
\end{tabular}
\label{metrics}
\end{center}
\end{table}

The evaluation of the proposed method on the test dataset was conducted by comparing classification model's appliance type predictions with ground-truth labels. Four most commonly used metrics in classification tasks were used to assess model performance: accuracy, macro precision, macro recall, and macro F1. The evaluation results, presented in Table~\ref{metrics}, indicate that the XGBoost model achieved the highest accuracy of 90\%, outperforming both the decision tree and logistic regression. Although decision tree and logistic regression models are inherently interpretable in case if features are not multi-collinear, their performance was consistently lower than the gradient boosting (XGBoost) model.

Additionally, for analyzing the efficiency of XGBoost in appliance classification, we plot confusion matrix, see Fig.~\ref{confusion_matrix}. The diagonal elements represent correctly classified activated/deactivated appliances, while off-diagonal elements indicate misclassifications. As shown in Fig.~\ref{confusion_matrix}, the model performs well for most appliances, except for soldering iron and light bulb. This can be explained by the fact that both devices are resistive and have similar current consumption magnitudes. Moreover, the confusion matrix reveals that other devices can be misclassified as well. For instance, air conditioner can be confused with fridge or fan, depending on their operating modes.

\section{Discussion}
\label{discussion}

An important consideration for the proposed framework is its latency, which is crucial for real-time applications. The latency ($\Delta T$) of the given approach is determined by the time required to record 19 (1 before and 18 after an event) cycles of the signal and by the time to execute each stage of the framework ($\tau$):

\begin{equation}
    \Delta T = \frac{19}{f_0} + \tau,
\end{equation}
where $f_0$ is the grid's frequency, e.g., 50 Hz or 60 Hz.

The computational complexity for each stage of the proposed framework is as follows:

\begin{itemize}
    \item FITPS executes in $O(T)$, where $T$ is the number of samples per cycle.
    \item z-score event detector has a constant runtime $O(w)$, where $w$ is the size of sliding window.
    \item Appliance signature estimation takes $O(T N_a \log T N_a)$.
    \item Fourier feature extraction has a complexity of $O(T \log T)$.
    \item XGBoost runs in $O(E\cdot D)$, where $E$ is the number of estimators, and $D$ is the maximum depth.
\end{itemize}

Thus, overall complexity of the proposed framework is $O(T N_a \log T N_a)$.

\section{Conclusion}
\label{conclusion}

In this paper, we proposed an explainable and interpretable framework for high-frequency event-based NILM. The method incorporates three key design principles: simplicity, computational efficiency, and explainability. Thus, it has the potential for real-time applicability, including deployment on edge devices. The framework was evaluated on the real-world PLAID dataset and reached 90\% appliance classification accuracy.

In future work, we aim to check the generalization ability of the framework across other datasets. Furthermore, we plan to test the method in real-world apartments with the high-precision real-time energy sensor.

\section*{Acknowledgment}

This research was conducted as part of the first author's industrial immersion project in collaboration with Monisensa Development LLC. The authors thank Monisensa for their guidance and for providing the computational resources for this research.

\bibliographystyle{IEEEtran}
\bibliography{ref}

\end{document}